\documentclass[fleqn,twoside]{article}
\usepackage{espcrc2}

\usepackage{graphicx}
\usepackage[figuresright]{rotating}

\newcommand{\AmS}{{\protect\the\textfont2
  A\kern-.1667em\lower.5ex\hbox{M}\kern-.125emS}}

\title{Dark halo densities, substructure, and the initial power spectrum}
\author{J. S. Bullock \address[osu]{
Department of Physics, The Ohio State University, 
Columbus, OH 43210, USA; \\
james@astronomy.ohio-state.edu; zentner@pacific.mps.ohio-state.edu}
\thanks{This work was supported by U.S. DOE Contract No.
DE-FG02-91ER40690.}
and  A. R. Zentner\addressmark[osu]
}
       
\begin{document}

\begin{abstract}

Although the currently favored cold dark 
matter plus cosmological constant model
for structure formation   assumes  an  $n=1$ scale-invariant
initial power  spectrum,  most  inflation models  produce at
least  mild deviations  from $n=1$.  Because  the lever arm
from the CMB normalization to galaxy scales is long, even  
a small ``tilt'' can have
important   implications  for galactic observations.  Here  we
calculate    the   COBE-normalized   power    spectra   for   several
well-motivated models of  inflation and compute implications for the
substructure content and central  densities of  galaxy halos.
Using an analytic model, normalized against N-body simulations, we show
that  while halos in  the standard  ($n=1$)  model are
overdense by a factor of $\sim 6$ compared to observations, several of
our  example inflation+LCDM models  predict halo densities well within
the range of  observations, which prefer models  with $n \sim 0.85$.
We go on to use a semi-analytic model  (also normalized against N-body
simulations) to follow the merger histories  of galaxy-sized halos and
track the orbital decay, disruption, and evolution of the merging
substructure. Models with $n
\sim 0.85$  predict  a factor of  $\sim 3$ fewer  subhalos at a
fixed circular velocity than the standard $n = 1$ case.  Although this
level  of reduction does   not resolve the ``dwarf  satellite
problem'',  it  does  imply  that   the level  of  feedback  required 
to match the observed number of dwarfs
is sensitive to   the initial power  spectrum.  Finally,  the fraction of
galaxy-halo mass that is bound up in substructure is 
consistent with limits imposed
by multiply  imaged quasars for all models considered: $f_{sat} > 0.01$
even for an effective tilt of $n \sim  0.8$.
 We conclude that, at their current level,
lensing constraints of this kind do not provide strong limits
on the primordial power spectrum.
\vspace{1pc}
\end{abstract}

\maketitle

\section{Introduction}

The  cold  dark matter  plus  cosmological  constant  (LCDM) model  of
structure formation is the  most successful and popular current theory
for  the  origin  of   universal  structure.   The  theory  is  highly
predictive, and given that observations constrain the universe to
be nearly flat  
(with $\Omega_m = 1 - \Omega_{\rm \Lambda}  \approx 0.3$, and $h
\approx  0.7$) a  crucial unknown  is the  primordial spectrum of
density  fluctuations.   It  is   usually  assumed  that  the  initial
fluctuation spectrum is scale-invariant, with $P(k) \propto k^n$, $n =
1$.   This choice is motivated by the fact that slow-roll inflation models 
predict {\em  nearly}  scale-invariant spectra.   However,
generic models  of inflation do  not predict primordial  power spectra
that  are {\em exactly}  scale-invariant and  almost all  models have
some small, often non-negligible,  ``tilt'' ($ n\ne1$) and ``running'' of
the spectral index ($ dn/d\ln k \ne 0$).  In this proceeding we report
on   our  work  to   model  the small-scale implications of  
reasonable  initial  power   spectra,  and  demonstrate  that  the
substructure  content and central  densities of  dark halos  should be
very sensitive to small deviations from scale-invariance.  Our models
are semi-analytic, but they are normalized and tested against N-body
simulations.  As we discuss below, our results may be important for 
interpreting galaxy rotation curves, dwarf galaxy counts, and 
substructure mass fractions as measured from multiply imaged quasars.

In Zentner \& Bullock \cite{zb} (hereafter ZB1 and ZB2)
we 
chose several well-motivated and representative inflation models 
and calculated the implied power spectra to second
order in slow-roll using the method of Stewart and Gong \cite{SG}.
Adopting $\Omega_{\rm  M} = 0.3$, $\Omega_{\rm
L} = 0.7$, $\Omega_{\rm b}h^2 = 0.020$, and $h = 0.72$ (LCDM),
we calculated P(k) using the results of Eisenstein \& Hu \cite{EH} and 
normalized to COBE on the pivot scale, $k_* \simeq 0.0023$ h$^{-1}$Mpc.  
In the interest of brevity, we report on only three models here:
\begin{itemize}
\item{IPL: $n = 0.94$, $dn/d\ln k = -0.001$, \\ $\sigma_8 = 0.83$}
\item{RM: $n = 0.84$, $dn/d\ln k = -0.004$, \\ $\sigma_8 = 0.65$}
\item{$n=1$: $n = 1.00$, $dn/d\ln k = 0.000$, \\ $\sigma_8 = 0.95$}.
\end{itemize}
The tilt and running are evaluated at $k=k_*$
and $\sigma_8$ is the rms fluctuation amplitude within spheres of
$8  h^{-1}$Mpc implied by the COBE normalization.
Observational constraints on $\sigma_8$ have historically favored  
a value near unity; however, some recent  estimates  indicate
surprisingly  low  values  of  $\sigma_8$  \cite{sig8}.
Roughly speaking,  $0.60 < \sigma_8 < 1.2$
spans the range of recently reported values 
(for $\Omega_m \simeq 0.3$).

\section{Central Densities}

\begin{figure}[htb]
\vspace{2pt}
\includegraphics[scale=0.35]{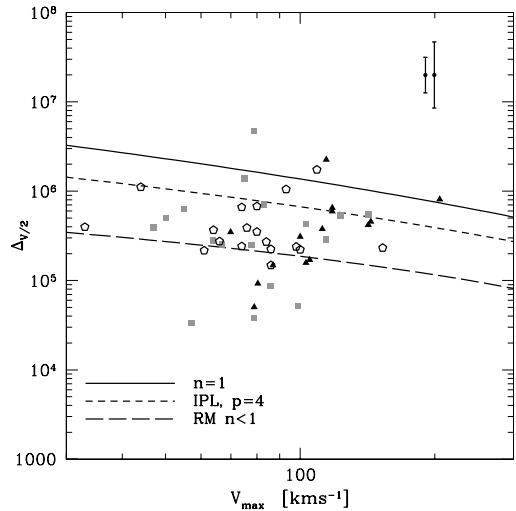}
\caption{The central density parameter (see text)
as a function of maximum circular velocity for several
dark matter-dominated galaxies (points).  The solid line corresponds
to the predicted densities for  standard $n=1$ LCDM halos, 
and the short and long-dashed
lines correspond to our expectations for two
different inflation-derived input power spectra.}
\label{fig:largenenough}
\end{figure}

The  theory of  CDM has  survived  remarkably well  since its creation
roughly twenty years ago  \cite{cdm}, but lingering for
nearly half this time has been a possible  problem with galaxy
central densities \cite{first}, which 
seem to be much less dense than the standard CDM model typically
predicts.  One part of this problem  --- the cusp vs. core
question --- concerns the precise inner slope  of galaxy rotation curves
and  whether they match the ``cuspy''  predictions of CDM theory. 
This particular question is highly debated, and  some argue that
observational errors and/or theoretical uncertainties make the
comparisons inconclusive.  But even among those who
believe there may be no problem  with the {\em shapes} of rotation
curves (the authors included), 
it is almost universally agreed upon that the data do prefer halos
that  are   less centrally  concentrated  than  typical  halos  in the
standard ($n=1$) LCDM model.

One way to see the problem is by comparing rotation curve data to
theory using the following, non-parametric
measure of the density \cite{alam}:
\begin{equation}
\label{eq:dv2def}
\Delta_{\rm V/2} \equiv \frac{\overline \rho (r_{\rm V/2})}{\rho_{\rm crit}} = 
\frac{1}{2}\Bigg(\frac{V_{\rm max}}{H_0 r_{\rm V/2}}\Bigg)^2.
\end{equation}
This  is the mean dark matter  density within the radius $r_{\rm V/2}$
where the galaxy  rotation curve falls to  half its maximum, 
$V_{\rm  max}$.  In practical units,
$\Delta_{\rm V/2}  \simeq    5 \times    10^5  (V_{\rm max}/100 \textrm{    kms}^{-1})^2 (r_{\rm V/2}/\textrm{h}^{-1}\textrm{   kpc})^{-2}$.

In Figure 1, we  plot $\Delta_{V/2}$ measured from several dark 
matter-dominated galaxy rotation curves (data from \cite{data}, 
see ZB1 for details) and compare to
predictions   for {\em typical} halo   densities in our models with 
different power spectra (lines).   Notice that the  $n=1$ model
overpredicts the galactic densities by a factor of $\sim 6$, while our
inflation-motivated models with $n < 1$ do much better.

The predictions were made using the semi-analytic  model of Bullock et  al.
\cite{b01}, which  was shown to reproduce the  halo densities measured
in N-body simulations  for $n=1$ and  $0.9$ LCDM power spectra as well
as a   variety of    power-law models   (see  ZB1   for  an  extensive
discussion).  The model  is   similar to that originally   proposed in
\cite{nfw}  (see also \cite{wechsler}), although  modified in order to
reproduce the redshift dependence seen in N-body halos. Qualitatively, it 
embodies the fact that halo densities reflect the density of the universe at 
their characteristic collapse redshift.    Halos   in  models   with   less
small-scale power collapse later, when the  universe is less dense, and therefore 
have   more diffuse cores.  Interestingly, even  the ``IPL''  case, with only 
a moderate tilt and mild running,  does significantly better than the 
$n=1$ standard model.  If we adopt the criterion that a "good" fit to 
the data has $\Delta_{V/2} \simeq 3 \times 10^5$ at $V_{\rm max} = 100$ kms$^{-1}$ 
then the data prefer $n(k_*) + 6.96dn(k_*)/d\ln k \simeq 0.85$, assuming that 
the primordial spectrum is the only element of the solution.

\section{Substructure}

As mentioned briefly in the previous section, the input power spectrum
greatly  affects the typical merger history of galaxy-sized halos.
Halos formed  in models with less power   on galactic scales typically
form later,  and  accrete a larger  fraction of  their substructure at
later epochs.  Similarly, the ``subhalos'' that are accreted also form
later,    are  less dense,   and therefore are  more easily  tidally
disrupted.  In order to explore how different initial power spectra
affect the substructure content of halos quantitatively, we developed a
semi-analytic model that tracks halo accretion histories
using the extended Press-Schechter formalism \cite{SK} and follows subhalo
orbital evolution in a semi-analytic fashion. 

\begin{figure}[htb]
\vspace{0pt}
\includegraphics[scale=0.35]{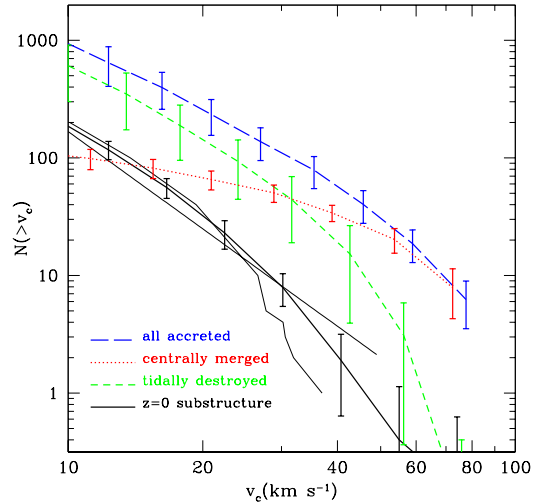}
\caption{Number of subhalos as a function of circular velocity.
The 
functions are derived for standard ($n=1$) LCDM host halos of
mass $M_{\rm vir}=10^{12} h^{-1} M_{\odot}$.  The long-dashed line
shows all accreted subhalos.  The short-dashed
line shows subhalos that were tidally destroyed, and
the dotted line those that merged with the central
object.  Subhalos that exist at $z=0$ are shown by the thick
solid line.  Error bars correspond to the rms scatter.
Subhalo counts from two different N-body 
simulations are shown by the thin solid lines.  Both agree reasonably
well with our model expectations.  }
\label{fig:acchist}
\end{figure}

\begin{figure}[htb]
\vspace{2pt}
\includegraphics[scale=0.35]{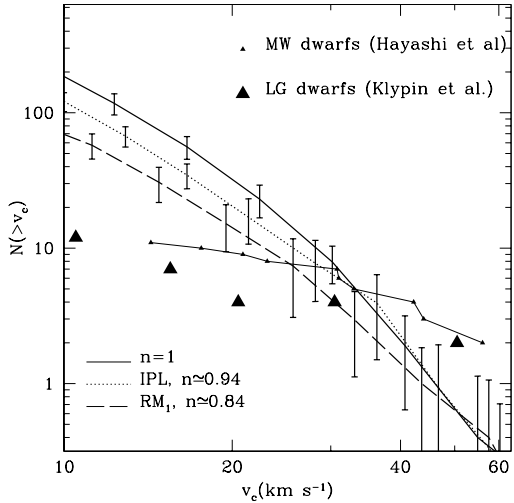}
\caption{The final subhalo abundances for our three models  compared
to two estimates of the local population of dwarf galaxies (points).}
\label{fig:subcount}
\end{figure}

Described  in  detail in ZB2, our   model  expands on the calculations
presented  in  Bullock, Kravtsov \& Weinberg  \cite{bkw}  and includes
some modifications similar to those advocated  
by  Taylor and Babul \cite{tb}.  Briefly,
after each merger event, we follow the subhalo  orbital evolution as it
decays  due to dynamical friction and loses mass via
tidal interactions with the host halo.  Dynamical friction dominates for large
subhalos, which typically merge with the central object soon after accretion.
Low mass substructures are efficiently stripped by the host potential,
and a large fraction of them become completely tidally destroyed.  The
results of our model for an ensemble of Milky Way-like host halos in
an $n=1$ LCDM cosmology are displayed in Figure  2.  Shown as a function of
$V_{\rm max}$  are the total number  of objects  that have  been
accreted compared to  the  number  that eventually
merge with  the   central object, the number  that are tidally destroyed, 
and the number that survive as distinct substructures at $z=0$.  It is  
clear that our model for the number of surviving structures is in 
good agreement with the results of N-body simulations (thin  solid lines) 
\cite{klyp}.  We use this model to estimate the 
sensitivity of the substructure population at $z=0$ to the 
primordial power spectrum.

\begin{figure}[htb]
\vspace{2pt}
\includegraphics[scale=0.35]{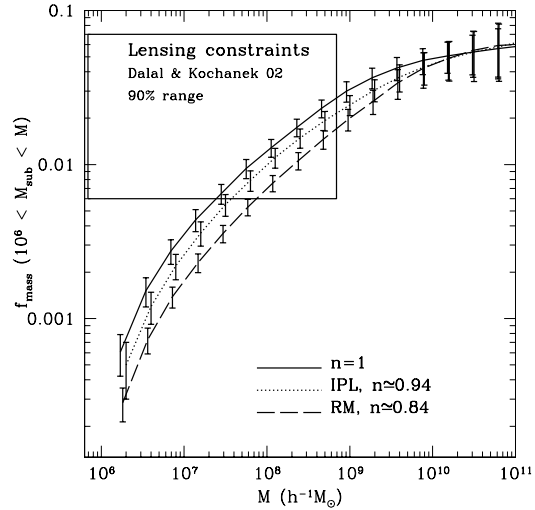}
\caption{The fraction of mass in a galaxy-host halo that is
contained in substructure less massive than $M$.
The box  roughly 
represents the lensing constraints 
of Dalal and Kochanek\cite{metcalf}. 
}
\label{fig:largenenough}
\end{figure}

Shown in Figure  3 are the cumulative  subhalo velocity functions for
each of our models compared to two different estimates of dwarf galaxy
counts   within $\sim    200$kpc   of  the     Milky   Way  and    M31
\cite{klyp,hayashi}.  That $n=1$ LCDM overpredicts the  number of
small satellites  by  roughly  an   order of magnitude is
expected.  This is the    well-known   dwarf    satellite    problem
(e.g. \cite{klyp}).  Interestingly, the number of subhalos
is reduced by as much as a  factor of $\sim 3$ for the inflation-derived
power spectra that we consider.  Although all  of  these  cases would
require some degree of feedback to match dwarf
counts (e.g., ionization \cite{bkw}), the {\em amount} of feedback  needed 
will depend sensitively on the input power spectrum.

Related to the  dwarf satellite problem  are recent efforts to measure
the substructure content  in halos using the  flux ratios  in multiply
imaged quasars   \cite{metcalf}.  For  example, Dalal and  Kochanek
used this  technique to constrain the fraction  of  the host halo mass
bound up in  substructure to be $0.006 \le  f_{\rm
 sat} \le 0.07$ (90
\% confidence).  They went on to  use this result to  try to limit the
tilt of   the primordial spectrum  using  a simplified 
model for the abundance of substructure and obtained $n \ge 0.94$ 
at 95\% confidence.  Our results on substructure
differ from those  of Dalal and Kochanek,  as illustrated in Figure 4.
We find that for a  host halo of  the relevant
mass, the total mass fraction in subhalos is typically larger than the
their lower limit ($f_{\rm sat}  \ge 0.006$) even for
our most tilted primordial  spectrum, $n \sim  0.84$. 
Since this model is on  the edge of what  is  acceptable by  $\sigma_8$ constraints, it
seems that strong lensing does not yet serve as a significant probe
of the primordial power spectrum.
(Using a  version of  our model, Dalal and   Kochanek are in  the
process of revising their  limits, and now  find results more in line with
those reported here \cite{dalal}.)


\newpage

\begin{thebibliography}{9}

\bibitem{alam} Alam, 
S.~M.~K., Bullock, J.~S., \& Weinberg, D.~H.\ 2002, ApJ, 572, 34 


\bibitem{b01} Bullock, J.~S., Kolatt, et al. \ 2001, MNRAS, 321, 559 

\bibitem{bkw} Bullock, J.~S., Kravtsov, A.~V., \& Weinberg, D.~H.\ 2000, ApJ, 539, 517 

\bibitem{dalal}
Dalal, N., 2002, private communication

\bibitem{EH} Eisenstein, D. J. and Hu, W. 1999, ApJ, 511, 5

\bibitem{first}
Flores, R. \&  Primack, J.R., 1994, ApJL, 427, L1; 
Moore, B. 1994,  Nature, 370, 629

\bibitem{hayashi} Hayashi, E., Navarro, J.F. et al. ApJ, submitted, astro-ph/0203004 

\bibitem{klyp} 
Klypin, A., Kravtsov, A.~V. et al. \ 1999, ApJ, 522, 
82; Font,  A.~S., Navarro, J.~F. et al.  2001, ApJL,  563, L1 



\bibitem{sig8} Melchiorri, A. \& Silk, J.  astro-ph/0203200; 
Seljak, U., astro-ph/0111362; Viana, P. T. P.,  Nichol,  R. C. \& 
Liddle, A. R., astro-ph/0111394; 
Bahcall, N.A.  et al., astro-ph/0205490.

\bibitem{metcalf} Metcalf, R.~B.~\& 
Madau, P.\ 2001, ApJ,  563, 9 ;  Chiba, M. 2002, ApJ, 565, 17;
Dalal, N.~\& 
Kochanek, C.~S.\ 2002, ApJ, 572, 25;
 Dalal, N.~\& 
Kochanek, C.~S.\ 2002, PrD, submitted, astro-ph/0202290;
Moustakas, L.A. \& Metcalf, R.B. 2002, MNRAS, submitted,
astro-ph/0206176






\bibitem{nfw}
Navarro,J.F., Frenk,  C. S. \&  White,S.D.M., ApJ, 1997, 490, 493 


\bibitem{cdm} Peebles, P.~J.~E.\ 1982, 
ApJL, 263, L1 ;  Blumenthal, S. M. Faber, J. R. Primack, and M. J. Rees, 
1984, Nature 311, 517 

\bibitem{SK} Somerville, R. S. \& Kolatt, T. S. 1999, MNRAS, 305, 1

\bibitem{SG} Stewart, E. D. \& Gong, J. O. 2001, Phys. Lett. B, 510, 1

\bibitem{data} Swaters, R. A., PhD Thesis, University of Groningen (1999); 
de Blok, W. J. G., McGaugh, S. S., \& Rubin, V. C. 2001, AJ, 122, 2396; de 
Blok, W. J. G. \& Bosma, A. A. 2002, astro-ph/0201276

\bibitem{tb} Taylor, J.~E.~\& 
Babul, A.\ 2001, ApJ, 559, 716 

\bibitem{wechsler} Wechsler, R.H., Bullock, J.S.,  et al. 2001, ApJ
568, 52

\bibitem{zb} Zentner, A.~R.~\& 
Bullock, J.~S.\ 2002, PrD, accepted, astro-ph/0205216 (ZB1);
  Zentner, A.~R.~\& 
Bullock, J.~S.\ 2002, in preparation (ZB2).



\end{thebibliography}
\end{document}